Modelling catastrophic risk in international equity markets: An extreme value approach

**JOHN COTTER**
**University College Dublin**

Abstract:
This letter uses the Block Maxima Extreme Value approach to quantify catastrophic risk in international equity markets. Risk measures are generated from a set threshold of the distribution of returns that avoids the pitfall of using absolute returns for markets exhibiting diverging levels of risk. From an application to leading markets, the letter finds that the Nikkei is more prone to catastrophic risk than the FTSE and Dow Jones Indexes.

**Address for Correspondence:**
Dr. John Cotter,
Centre for Financial Markets,
Graduate School of Business,
University College Dublin,
Blackrock,
Co. Dublin,
Ireland.
E-mail. john.cotter@ucd.ie

April 6 2005

# Modelling catastrophic risk in international equity markets: An extreme value approach

I Introduction:

Tail returns can be catastrophic for investors and accurate modelling of these is paramount. This letter uses the Block Maxima Extreme Value approach to quantify catastrophic risk for investors with long and short positions in international equity markets.[1] Quantitatively, catastrophic risk occurs if market movements exceed some extreme threshold value. We fit the Generalised Extreme Value (GEV) distribution to leading equity indexes that only models the tail returns of a probability distribution associated with catastrophic risk.[2] Risk measures are generated from a set threshold of the distribution of returns (99% level) that avoids the pitfall of using absolute returns for markets exhibiting diverging risk.

Catastrophic events relate to extraordinary trading periods that cannot be reconciled with previous and subsequent market movements. Thus, these events distinctively belong to a separate distribution distinct from ordinary market movements and should be modelled separately. Extreme Value Theory (EVT) is an optimal approach to quantifying the extent of rare catastrophic events. First, and most important, EVT dominates alternative frameworks in modelling tail events (Longin, 2000, Cotter, 2004a). Second, event risk is explicitly taken into account by EVT since it explicitly focuses on extreme events. Third EVT reduces model risk since it does not assume a particular model for returns. Finally, it avoids coarseness and bias in the tail estimates and produces a useful risk language for promoting high-risk concepts.

---

[1] In contrast, in a qualitative sense, catastrophic risk requires market participants such as investors and bankers agreeing on the occurrence of extreme events. Kindleberger (2001) notes that these extreme events are a result of irrational speculation in the form of manias and panics, accurately describing the large decline in international markets during the 1987 crash.
[2] Alternative applications of EVT include modelling margin requirements (Dewachter and Gielens, 1999) and stability in foreign exchange markets (Cotter, 2005).

The letter is organised as follows: section II describes the risk measures and estimation procedure. Using three leading market indexes from different geographical regions, the Dow Jones Industrial Average, the Nikkei 225 and the FTSE All Share, section III follows with extreme value estimates and discussion.

II Risk measures and estimation procedure

Using techniques from EVT this study fits a GEV to the data using the Block Maxima approach that models the maxima, *M*, for the upper tail for some block of time, for example, a year.[3] Assuming a random variable *X*, for finite samples the following three parameter version of the GEV is as follows:

$$H_{\xi,\sigma,\mu} = H_\xi\left(\frac{x-\mu}{\sigma}\right) \quad x \in \mathsf{D}, \quad \mathsf{D} = \begin{cases} ]-\infty, \mu - \sigma/\xi[ & \xi < 0 \\ ]-\infty, \infty[ & \xi = 0 \\ ]\mu - \sigma/\xi, \infty[ & \xi > 0 \end{cases} \quad (1)$$

where $\mu$ is the location parameter, $\sigma$ is the scale parameter and $\xi$ is the shape parameter of the extreme value distribution. The shape parameter is the key to using EVT as it separates three types of extreme value distributions according to different shapes, the Gumbel ($\xi = 0$), Weibull ($\xi < 0$) and Fréchet ($\xi > 0$) distributions. The latter extreme value distribution is supported in the finance literature as it exhibits a fat-tails property, also found for market returns (Cotter and McKillop, 2000).

---

[3] EVT is commonly applied in the financial economics literature, and for a comprehensive discussion of the theoretical framework see Embrechts et al (1997). Following convention we will focus on the maxima but alternatively we could detail the minima for the lower tail of a distribution. Alternative non-pararmetric approaches have also been applied in modelling tail behavour (see Cotter, 2004b)

The maximum likelihood estimation procedure yields parameter estimates for $\xi$, $\mu$ and $\sigma$ by maximizing the following log-likelihood function with respect to the three unknown parameters

$$L(\xi,\mu,\sigma;x) = \prod_i \log(h(x_i)), \qquad x_i \in M$$

where

$$h(\xi,\mu,\sigma;x) = \frac{1}{\sigma}\left(1+\xi\frac{x-\mu}{\sigma}\right)^{-1/\xi-1} \exp\left(-\left(1+\xi\frac{x-\mu}{\sigma}\right)^{-1/\xi}\right) \qquad (2)$$

is the probability distribution function for $\xi \neq 0$ and $1+\xi\frac{x-\mu}{\sigma} > 0$. Confidence intervals for these parameters $\hat{\xi}$, $\hat{\mu}$ and $\hat{\sigma}$ can easily be obtained via the profile log likelihood function.

The GEV parameters are used to generate the catastrophic risk measures. Taking an extreme threshold or $q$th quantile of a continuous distribution with distribution function $F$ is

$$x_q = F^{-1}(q)$$

where $F^{-1}$ is the inverse of the distribution function. The GEV catastrophic risk level, $x_{n,k}$, using the maximum likelihood parameters is:

$$x_{n.k} = H^{-1}_{\xi,\mu,\sigma}(1-1/k) = \begin{cases} \hat{\mu} - \frac{\hat{\sigma}}{\hat{\xi}}\left(1-(-\log(1-1/k))^{-\hat{\xi}}\right) & \xi \neq 0 \\ \hat{\mu} - \hat{\sigma}\log(\log(1-1/k)) & \xi = 0 \end{cases} \qquad (3)$$

Again asymmetric confidence intervals for the catastrophic risk level can be calculated using the profile log likelihood function.

To illustrate, consider a model using daily returns and a block size corresponding to annual maxima ($\approx 261$ days). The $k$-year catastrophic risk level $x_{261,k}$ is defined as

$$P\{M_{261} > x_{261,k}\} = 1/k, \qquad k > 1 \qquad (4)$$

This is the return level that we expect to exceed only in one year out of every $k$ years, on average and has a probability $1/k$. Now we turn to our estimation and inferences.

III Results and discussion

The analysis is completed on daily logarithmic returns series from liquid US, Asian and European markets between January 1, 1985 and December 31. The indexes chosen are the well-known Dow Jones Industrial Average, the Nikkei 225 and the FTSE All Share. Findings from a representative selection of indexes are given.

In Table 1 summary statistics detailing the first four moments, min and max values and the Jarque-Bera normality test are given for the full distribution of returns and for a subset of values incorporating 10 percent of the full sample. The latter analysis is to investigate the tail behaviour of financial returns, as it this part of the distribution that gives rise to catastrophic risk and our application of GEV. Overall, we find the mean of upper and lower tail returns deviate substantially from the approximately zero mean of the full distribution of returns, with the Nikkei exhibiting the largest deviations. Moreover, daily risk is approximately 1% although some very large single day returns occur.

INSERT TABLE 1 HERE

Standard financial time series properties are recorded for the tail and full distributions namely, a lack of normality, due to excess skewness, and excess kurtosis. To investigate this latter property in more detail, Figure 1 presents QQ plots of quantiles of the observed distribution set against the normal distribution for both the full set of Dow returns and for the subsets of tail values. Two obvious points are clear. First, all distributions exhibit fat-tails. Second, the fat-tail characteristic becomes more pronounced for the tail returns. These plots drive our application of the GEV and in particular, the Fréchet GEV.

INSERT FIGURE 1 HERE

Turning to the extreme value analysis, maximum likelihood parameters of the fitted GEV to the upper tails of the indexes are given in table 2. The dispersion parameter values concur with the summary statistics of the tail distributions, indicating that the Nikkei index fluctuates more than its counterparts. And as expected, the location and dispersion estimates increase as interval size increases. As stated, the most important parameter for modelling and distinguishing tail behaviour is the shape parameter. We find all point estimates are positive, and generally there is support for the hypothesis of returns converging to the fat-tailed Fréchet distribution at a 95% confidence levels. Specifically, the fattest tail shape recorded is for the FTSE index at a quarterly interval with a shape point estimate of 0.361. Variation in tail shape does occur across markets, and interval of estimation, where no systematic pattern occurs. For example, the shape parameter is reasonably constant across the intervals for the Dow whereas it decreases for the Nikkei. This has implications for the modelling of catastrophic risk where each asset should be modelled separately, and for different frequencies.

INSERT TABLE 2 HERE

Taking the three EV parameters, we now estimate the catastrophic risk levels for a 99.9% confidence level and these are given in Table 3. This allows us to obtain information on the size and frequency of catastrophic risk levels. Here we show the estimated *20*-month, *20*-quarter and *20*-semester catastrophic risk levels for the upper and lower tail of each index. These have an attractive inference with for example, a *20-month* catastrophic risk level representing a level that we expect to exceed in *one month* out of every *twenty months* on average. So for example, the *20-month* catastrophic risk level for the upper tail of the Nikkei index is 5.82% implying that positive extreme price movements of this magnitude are expected in this market once every 20 months on average.

INSERT TABLE 3 HERE

Some interesting findings are noted. Catastrophic risk increases as you increase the interval size where investors would experience larger absolute returns from these major markets. Furthermore, with the exception of the Dow for monthly blocks, lower and upper catastrophic risk is similar and is within the respective confidence intervals for each interval block. However in terms of identifying the riskiest market at this interval, we find that the Nikkei exhibits the largest levels, and the FTSE exhibits the smallest levels, of extreme returns.

An overall portfolio return is driven by these extreme catastrophic values as investor performance is frequently the end result of a few exceptional trading days as most of the other days only contribute marginally to the bottom line. Hence correct modelling of catastrophic risk is vital.

Acknowledgements: University College Dublin's Faculty research funding is gratefully acknowledged.

Full series

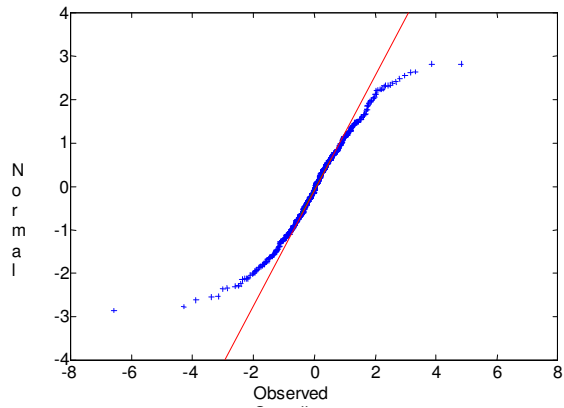

Upper tail

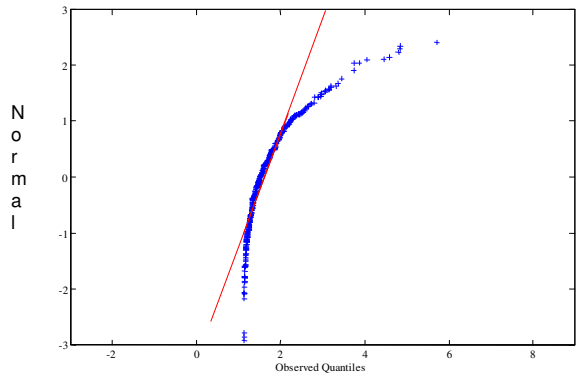

Lower tail

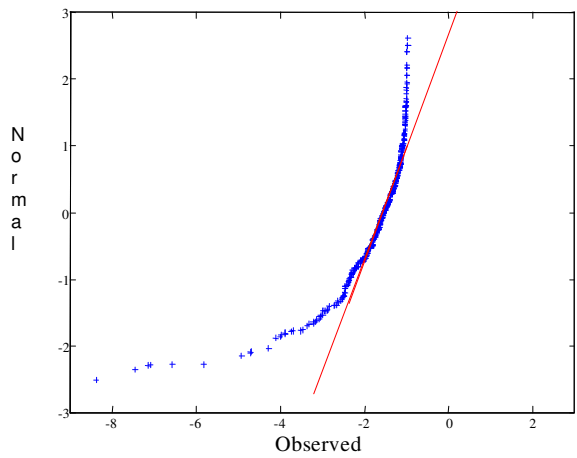

Figure 3. Q-Q plots of Dow Jones Industrial Average returns.
This figure plots the quantiles of the observed distribution against the normal distribution (straight line) for the full series and upper and lower 10 percent of returns.

Table 1. Summary statistics for daily index series

| Index | Mean | Std D | Min | Max | Skew | Kurt | J-B |
|---|---|---|---|---|---|---|---|
| **DOW** | 0.0524 | 1.06 | -25.64 | 9.67 | -3.74 | 92.32 | 1397123 |
| Upper | 1.76 | 0.76 | 1.12 | 9.67 | 4.14 | 33.55 | 17404 |
| Lower | -1.80 | 1.50 | -25.64 | -0.97 | -10.36 | 154.85 | 408088 |
| **FTSEALL** | 0.0387 | 0.87 | -11.91 | 5.70 | -1.31 | 20.32 | 53380 |
| Upper | 1.45 | 0.59 | 0.98 | 5.70 | 3.28 | 18.83 | 5104 |
| Lower | -1.53 | 0.96 | -11.91 | -0.91 | -6.04 | 55.85 | 51071 |
| **NIKKEI** | 0.0043 | 1.34 | -16.14 | 12.43 | -0.17 | 13.01 | 17453 |
| Upper | 2.39 | 1.17 | 1.40 | 12.43 | 3.23 | 20.12 | 5820 |
| Lower | -2.48 | 1.15 | -16.14 | -1.45 | -4.93 | 50.80 | 41392 |

Notes: The summary statistics are presented for each index as well as the upper and lower 10 percent of realisations. Mean, min, max, standard deviation (Std D) values are presented in percentages. Normality is formally examined with the Jarque-Bera (J-B) test which a critical value of 3.84. All the skewness (Skew), kurtosis (Kurt) and normality coefficients are significant at the 5 percent level.

Table 2. Parameter estimates for upper tail of index series

| Index | Block Length | $\hat{\xi}$ | $\hat{\sigma}$ | $\hat{\mu}$ |
|---|---|---|---|---|
| **DOW** | Month | 0.167 [0.075, 0.271] | 0.546 [0.495, 0.609] | 1.411 [1.339, 1.478] |
|  | Quarter | 0.168 [0.025, 0.354] | 0.716 [0.607, 0.869] | 1.908 [1.739, 2.057] |
|  | Semester | 0.170 [-0.002, 0.382] | 0.825 [0.665, 1.091] | 2.241 [1.959, 2.477] |
| **FTSEALL** | Month | 0.259 [0.150, 0.373] | 0.388 [0.352, 0.432] | 1.107 [1.059, 1.151] |
|  | Quarter | 0.361 [0.147, 0.586] | 0.456 [0.389, 0.555] | 1.434 [1.340, 1.513] |
|  | Semester | 0.214 [-0.007, 0.510] | 0.644 [0.514, 0.859] | 1.800 [1.585, 1.978] |
| **NIKKEI** | Month | 0.294 [0.172, 0.423] | 0.872 [0.791, 0.973] | 1.688 [1.582, 1.784] |
|  | Quarter | 0.114 [-0.042, 0.337] | 1.407 [1.184, 1.715] | 2.696 [2.354, 3.003] |
|  | Semester | 0.052 [-0.231, 0.301] | 1.765 [1.392, 2.351] | 3.434 [2.795, 3.993] |

Notes: Extreme value parameters, the tail index ($\xi$), the scale parameter ($\sigma$), and the location parameter ($\mu$) are estimated via maximum likelihood methods. 95% confidence intervals are given in []. Block lengths of 192 extremes (month), 64 extremes (quarter) and 32 extremes (semester) are used.

Table 3. Estimated catastrophic risk levels of index series

| Index | Tail | Month | Quarter | Semester |
|---|---|---|---|---|
| DOW | Lower | 4.32 [3.71, 5.29] | 6.47 [4.90, 10.09] | 9.04 [6.05, 19.40] |
|  | Upper | 3.51 [3.14, 4.07] | 4.66 [3.92, 6.25] | 5.45 [4.33, 8.40] |
| FTSE | Lower | 3.21 [2.80, 3.87] | 4.07 [3.29, 5.97] | 5.02 [3.63, 10.69] |
|  | Upper | 2.84 [2.50, 3.39] | 3.89 [3.03, 6.04] | 4.47 [3.49, 7.87] |
| NIKKEI | Lower | 5.46 [4.77, 6.60] | 7.09 [5.85, 9.71] | 8.81 [6.62, 16.69] |
|  | Upper | 5.82 [4.95, 7.28] | 7.67 [6.38, 10.65] | 9.10 [7.33, 14.59] |

Notes: The table shows the estimated *20*-month, *20*-quarter and *20*-semester catastrophic risk levels for the upper and lower tail of each index. 95% confidence intervals are given in [].